\documentclass[letterpaper,10pt,conference]{ieeeconf}
 \usepackage{physics}
\IEEEoverridecommandlockouts      
\author{Shuang Gao %
\thanks{Shuang Gao is with the Department of Electrical and Computer Engineering, and Centre for Intelligent Machines, McGill University,
   Montreal, QC, Canada, H3A 0E9. 
         {Email: \tt\small    $\{$sgao$\}$@cim.mcgill.ca}.  }}

\usepackage[noadjust]{cite}

\usepackage{graphics}
\usepackage{epstopdf}

\usepackage{amsmath}
  \usepackage[font=footnotesize]{subfig}

\usepackage{enumerate}
\usepackage{color}

\newcommand*\TRANS{{\mathpalette\doTRANS\empty}}
\makeatletter
\newcommand*\doTRANS[2]{\raisebox{\depth}{$\m@th#1\intercal$}}
\makeatother

\usepackage{url}
\usepackage{amssymb,mathtools}

\usepackage[standard,thmmarks,amsmath]{ntheorem}
\usepackage{times}
\usepackage{tikz}

\DeclareMathOperator{\diag}{\textup{diag}}
\usepackage[export]{adjustbox}

\begin{document}
%
\title{\huge  Centrality-Weighted Opinion Dynamics:\\ Disagreement and Social Network Partition}
%
%
%

\maketitle 
\begin{abstract}
This paper proposes a network model of opinion dynamics based on both the social network structure and network centralities.  
The conceptual novelty in this model is that the opinion of each individual is weighted by the associated network centrality in characterizing the opinion spread on social networks. 
Following a degree-centrality-weighted opinion dynamics model, we provide an algorithm to partition nodes of any graph into two and multiple clusters based on opinion disagreements. 
Furthermore, the partition algorithm is applied to real-world social networks including the Zachary karate club network \cite{zachary1977information} and the southern woman network \cite{davis1941deep} and these application examples indirectly verify the effectiveness of the degree-centrality-weighted opinion dynamics model. Finally, properties of general centrality-weighted opinion dynamics model  are established.
%
\end{abstract}

\section{Introduction}
\subsection{Motivation and background}
Posts on online social platforms, opinions of individuals in social networks, or ideas in research papers, tend to be weighted (or perceived) differently due to the heterogeneity of hyperlink networks, social connections, or citation structures. These differences may be influenced by the ranking in Google search, social importance in social networks, or citation counts in citation networks. 
Network centralities, which quantify how central nodes are in a network, play a natural role in the phenomenon of heterogeneous weights (in posts, opinions or ideas) in these examples above. 
%
In particular, centrality weights proportional to  nodal connection degrees may appear naturally in networks (see  for example the preferential attachment network-growth model \cite{barabasi1999emergence} for scale-free networks).  
Network influence weighted by connection degrees may well represent  an underlying natural principle in the evolution of opinions and influences on social networks.

The modelling of opinion dynamics dates back to the work of French \cite{french1956formal} via agent-based models on directed graphs,  the work of Degroot \cite{degroot1974reaching} based on Markov process and that of Friedkin and Johnsen \cite{friedkin1990social} based on dynamical systems.  There have been many useful variations based on these basic models (see for instance \cite{anderson2019recent, hegselmann2002opinion,altafini2012consensus,jia2015opinion,xu2015modified,proskurnikov2017tutorial} and the references therein for an overview). 
The study of opinion dynamics \cite{french1956formal,degroot1974reaching,friedkin1990social} connects inherently  to  the study of distributed coordination or the consensus protocol (see for instance
\cite{saber2003consensus,jadbabaie2003coordination,lin2004local}), which employs locally the relative state differences among neighbours. 
 In this type of modelling, it is typically assumed that the ``influence matrix'' or ``influence network'' is given beforehand.  However the influence matrix is not necessarily the social network structure and to the best knowledge of the author, there lacks a systematic way of identifying the influence matrix from network structures in the study of opinion dynamics. 
A second type of models to study the mechanism of opinion evolution and  decision-making on a social network is via the Ising model with state-dependent network interactions \cite{sznajd2000opinion}, and  the large-scale network opinion analysis is based on statistical characterizations of the underlying opinions \cite{toscani2006kinetic,albi2016opinion}. For this type of models, it is quite difficult to obtain explicit solutions and one typically needs to rely on numerical simulations or mean-field approximations \cite{sznajd2000opinion,toscani2006kinetic,albi2016opinion}. 
In addition, plenty of research papers have approached the modelling of opinion evolution from Bayesian update perspectives (see for instance, \cite{acemoglu2011opinion,orlean1995bayesian,rahimian2016bayesian,salhab2020social}). 
Readers are referred to \cite{anderson2019recent,proskurnikov2017tutorial} for an overview of the extensive research in modelling opinion dynamics.
In the current paper, the basic model of studying opinion dynamics over networks is based on \cite{degroot1974reaching, friedkin1990social}

Centralities have been employed in modelling opinion dynamics in various different ways.
The work \cite{kandiah2012pagerank} studied the opinion formation problem by incorporating the Ising model with the page-rank centrality \cite{brin1998anatomy}. The paper \cite{singh2019centrality} provided a dynamics update model based on Erd\"os-R\'enyi random graphs and  centralities such as in-degree, closeness and page-rank centralities. A centrality notion as the asymptotic opinion state was studied in \cite{jia2015opinion}. These papers, among others, are different from the current paper in both the model formulation and the use of centralities in modelling opinion dynamics.

Social choice problems based on network structures are essentially graph partition problems, which have been studied from various perspectives (see for instance \cite{fiedler1973algebraic,pothen1990partitioning,spielman1996spectral,newman2006modularity,white2005spectral}). Such problems also arise from image segmentation (see for instance \cite{peng2013survey} the references therein). 
Various approaches, such as minimal spanning tree methods \cite{peng2013survey}, spectral partition \cite{fiedler1973algebraic,spielman1996spectral}, modularity matrix approach \cite{newman2004finding,newman2006modularity}, have been used to solve graph partition problems (see \cite{peng2013survey,hoffman2018detecting} for a survey of  different methods).
In particular, 
spectral partition forms an important heuristic method for partitioning graphs \cite{spielman1996spectral}. 
The study of graph spectral partition method started in 1970s in   \cite{donath1972algorithms,donath1973lower} based on eigenvectors of adjacency matrices and  in \cite{fiedler1975eigenvectors,fiedler1975property} based on the Fielder eigenvector (i.e., the eigenvector associated with the smallest non-zero eigenvalue of the graph Laplacian). 
In this paper, a spectral partition based on the centrality-weighted network provides us an algorithm to partition social networks. 

\subsection{Contribution}
We propose a simple and concrete procedure to build ``influence networks'' based on social network structures and network centralities for the modelling of opinion dynamics. %
%
Then following a degree-centrality-weighted opinion dynamics, we provide a method for network partitions based on the disagreements approximately represented by the projection of the opinion state in the most significant eigendirection that is orthogonal to agreement subspace. This partition method produces the exact result for the  split of the Zachary's karate club network \cite{zachary1977information}, which indirectly verifies the effectiveness of our degree-centrality-weighted opinion dynamics model. Compared to network partition algorithm in \cite{newman2006modularity} based on modularity measure which also correctly characterizes the partition for  the Zachary's karate club network \cite{zachary1977information}, our method  provides  a different theoretical interpretation of the network partition based on the disagreement of opinions  from a dynamical system point of view,  and this explanation fits naturally into the context of the social choice problem in~\cite{zachary1977information}.

\subsection*{Notation and terminology}
Let  $[n]: = \{1,...,n\}$. 
Let $\text{diag}(u)$ represent the diagonal matrix with the diagonal elements specified by the vector $u$. $\text{diag}^{\frac{1}{2}}(u)$ (resp. $\text{diag}^{-\frac{1}{2}}(u)$) denotes the diagonal matrix with diagonal elements given by $u(i)^{\frac1{2}}, i \in [n]$ (resp. $u(i)^{-\frac1{2}}, i \in [n]$). $\mathbf{1}$  denotes the $n$-dimensional column vector with all elements being one.  
For any matrix $A$, $A^\TRANS$ to denote its transpose.  

A square matrix $A$ is \emph{diagonalizable} if there exist an invertible real matrix $P$ and a  diagonal real matrix $\Lambda$ such that
$A = P^{-1} \Lambda P$. An \emph{eigenvalue} $\lambda$ of a matrix $A$  is defined as the complex or real number such that there exists a complex or real vector $v\neq 0$ satisfying
$Av= \lambda v$. Then $v$ is called an \emph{eigenvector} of $A$ associated with the eigenvalue $\lambda$.  Let $I$ denote the identity matrix with an appropriate dimension. Then the \emph{algebraic multiplicity} of an eigenvalue $\lambda$ of $A$
 is defined as the multiplicity of  $\lambda$ as a root of $\text{det}(\lambda I - A)$; the \emph{geometric multiplicity} of an eigenvalue $\lambda$ is defined as the maximum number of linearly independent eigenvectors associated with $\lambda$, which can be computed by $n-\text{rank}(\lambda I -A)$ when $A$ is of dimension $n\times n$. 
 In this paper, whenever we list the eigenvalues $\{\lambda_1,\cdots, \lambda_n \}$ of a matrix, we always order the eigenvalues in a non-decreasing order, that is, 
$
\lambda_1\leq \lambda_2\leq \cdots \leq \lambda_n.
$

A graph $G=(V,E)$ is defined by the set of nodes $V$ and the set of edges $E\subset V \times V$ connecting the nodes. A weighted graph is a graph in which each edge is associated with a real number weight. Taking the node set $[n]$, then a graph can be represented by its adjacency matrix $A=[a_{ij}]$ where $a_{ij}$ denotes the weight from node $j$ to node $i$.  A graph is \emph{undirected} if the connection weight from $i$ to $j$ is always equal to the weights from $j$ to $i$ for all $i,j \in [n]$ (that is, its adjacency matrix $A$ satisfies $A= A^\TRANS$).  $\mathcal{G}(A)$ denotes the graph with $A$ as the adjacency matrix and $[n]$ as the node set. $N_i:=\{j| (j,i) 
\in E\}$ denotes the (incoming) neighborhood of node $i$.
 A \emph{path} (resp. \emph{directed path}) from node $i$ to $j$ is defined as a finite or infinite sequence of edges which joins a sequence of distinct nodes (resp. and the directed edges are in the same direction).
An undirected graph is \emph{connected} if for every node pair on the graph one can identify a path. A directed graph is \emph{strongly connected} if every node pair $(i,j)$ on the graph has a directed path from $i$ to $j$.

\section{Opinion evolution over social networks}
\subsection{Degree-centrality-weighted opinion dynamics}
Social conformity characterizes the phenomenon that individuals tends to change his/her behavior or response to conform with that of the group and it has supporting evidence in social and psychology studies  (see for instance \cite{cialdini2004social}). Considering social conformity, 
 basic assumptions in our degree-centrality-weighted opinion dynamics model are as follows: 
\begin{enumerate}
\item[(i)] Individuals on a social network communicate and change their own opinions in the direction to conform with those of their neighbours; 
\item[(ii)] Each individual weights these influences from the neighbours based on their importance on the network in terms of the number of their connections. 
\end{enumerate}
Based on these basic assumptions, the dynamics for opinion evolution over a social network are then formulated as follows:\\
\begin{equation}\label{eq:opinion-evo}
 	\tau \dot {x}_i =  \sum_{j\in N_i}\frac{d_j}{\sum_{j\in N_i} a_{ij}d_j} a_{ij}(x_j-x_i), ~~ x_i(0)= x_{i0}, ~ i \in [n]
 \end{equation}
where $d_i = \sum_{j\in N_i} a_{ij}$ is the degree (centrality) of node $i$ on the network, $a_{ij}$ is social connection weight between nodes $i$ and $j$ (which can take a real value  if the underlying social network is weighted, or  0-1 value if the underlying graph only characterizes the social network structure), and $\tau>0$ is a fixed time constant. %

Let $A=[a_{ij}]$ denote the adjacency matrix of the underlying graph.  The degree centrality vector is given by
$
d = A \mathbf{1}, ~  \text{i.e.,} ~  d_i = \sum_{j\in N_i} a_{ij},  ~ i \in [n].
$
Then the ``influence matrix'' for \eqref{eq:opinion-evo} is given by 
$$
\bar A  = [\textup{diag}(h)]^{-1} A  \textup{diag}(d), \quad\text{with}~h = A d.
$$ 
Clearly $\bar A$ is not necessarily symmetric. 
The corresponding  Laplacian matrix  of this weighted ``influence matrix'' $\bar {A}$ is then given by
$$
\bar {L}  \triangleq \textup{diag}(\bar A \mathbf{1}) -\bar A = I_n - \bar A,
$$
where the last equality holds because 
\begin{equation} \label{eq:property-Abar}
\bar A \mathbf{1} = [\textup{diag}(h)]^{-1} A  \textup{diag}(d) \mathbf{1} = [\textup{diag}(Ad)]^{-1} A d =\mathbf{1}.
\end{equation}
We note that the Laplacian matrix $\bar L$ is different from normalized Laplacian matrices  (by connection degrees) of $A$  which are typically given by 
\[
\begin{aligned}
 &L_{n} \triangleq [\textup{diag}(d)]^{-1} (\textup{diag}(d)- A) =
I_n -[\textup{diag}(d)]^{-1}  A,\\
&L_{sn}\triangleq 
I_n -[\textup{diag}(d)]^{-\frac12}  A [\textup{diag}(d)]^{-\frac12}.
\end{aligned}
\]

Denote $x = [x_1,\ldots, x_n]^\TRANS$. Then the dynamics in \eqref{eq:opinion-evo} have the following  compact representation
\begin{equation}\label{eq:compact-dynamics}
	 \tau  \dot{x} = - \bar{L} x, \quad x(0) = x_0
\end{equation}
where the Laplacian matrix $\bar L$ and the adjacency matrix $\bar A$  are respectively given by $$\bar{L} = I_n - \bar A, \quad \text{with}~ \bar A  = [\textup{diag}(Ad)]^{-1} A  \textup{diag}(d).$$
\subsection{Properties of the influence matrix $\bar A$ and its Laplacian $\bar L$}
\begin{proposition}[Appendix \ref{app:prop1}] \label{prop:one-zero-eigenvalue}
If the underlying graph $\mathcal{G}(A)$ with the adjacency matrix $A$ is connected and undirected, then the Laplacian matrix $\bar L$ contains only one zero eigenvalue and all the other eigenvalues of $\bar L$ have strictly positive real parts. 
\end{proposition}

If the underlying graph $\mathcal{G}(A)$ with the adjacency matrix $A$ is connected and undirected, then Proposition \ref{prop:one-zero-eigenvalue} implies that
 the long-term behaviour of the system model \eqref{eq:compact-dynamics} reaches agreement as follows
  $
 \lim_{t\rightarrow \infty}x(t) =    u_1 (v_1^\TRANS x_0) 
$
where $u_1 =\frac{1}{\sqrt{n}}\mathbf{1}$  is the normalized right eigenvector of $\bar L$ and $v_1$ is the normalized left eigenvector of $\bar L$ that associated with the only zero eigenvalue $\lambda_1=0$. (See for instance \cite{mesbahi2010graph}). 

\begin{proposition}[Appendix \ref{app:prop2}]\label{prop:realeigenpairs}
	If the underlying graph $\mathcal{G}(A)$ with the adjacency matrix $A$ is connected and undirected, then all the eigenvalues of $\bar A$ and  $\bar L$ are real. 
\end{proposition}

The conclusion in Proposition \ref{prop:realeigenpairs} on real eigenvalues for $\bar A$ and $\bar L$ may not hold when $A$ is not a symmetric matrix.  

\begin{proposition}[Appendix \ref{app:prop3}]\label{prop:diagonalizable}
If the underlying graph $\mathcal{G}(A)$ with the adjacency matrix $A$ is connected and undirected, then $\bar{A}$ and
$\bar L$ are diagonalizable. 	
\end{proposition}
%
Henceforth we assume that $\mathcal{G}(A)$ with the adjacency matrix $A$ is undirected and connected.  Based on Proposition \ref{prop:diagonalizable}, $\bar L$ is diagonalizable,  and hence
 the solution to \eqref{eq:opinion-evo}  is explicit  given by 
\begin{equation}\label{eq:explicit-evo-solution}
	x(t) = \sum_{i=1}^n e^{-\frac{t}{\tau}\lambda_i }   u_i  (v_i^\TRANS x_0),
\end{equation}
where $v_i^\TRANS$ and $u_i$ represent respectively the left and right orthonormal eigenvectors of $\bar L$ associated with eigenvalue  $\lambda_i$ (allowing repeated eigenvalues),  $i\in[n]$.

\subsection{Network opinion partition algorithm}

The \emph{disagreement state} is defined as the state projection into the subspace that is orthogonal to the agreement subspace $\textup{span}\{\mathbf{1}\}$ (which is equivalently $\textup{span}\{u_{1}\}$) in the opinion state space. Then the disagreement state of \eqref{eq:explicit-evo-solution} is given by 
\begin{equation}\label{eq:disgreement}
	x^\textup{dis}(t) = \sum_{i=2}^n    u_i  (v_i^\TRANS x(t))= 
 \sum_{i=2}^n e^{-\frac{t}{\tau}\lambda_i }   u_i  (v_i^\TRANS x_0).
\end{equation}
 The disagreement state converges to the origin exponentially over time and the slowest rate is governed by the smallest non-zero eigenvalue  $\lambda_2(\bar L)$ of $\bar L$. Hence to approximately characterize the disagreement over networks, one may use the smallest non-zero eigenvalues $\lambda_2(\bar L)$ and its associated left and right eigenvectors. 
 
 The basic idea for the partition algorithm is then to analyze the opinion state projected into the subspace associated with the eigenvector(s) of $\lambda_2(\bar L)$.  
  The signs of this projected opinion state will cluster nodes into two groups.  \vspace{5pt}\\
\textbf{Partition Algorithm (Social Choice Algorithm):}
\begin{enumerate}[\bf (S1)]
 \item When  $\lambda_2(\bar L)$ has algebraic multiplicity~1, let 
\[
s \triangleq  u_2 .
\]

 When $\lambda_2(\bar L)$  has algebraic multiplicity $m_2~ (m_2\geq 2) $, let  
\begin{equation}\label{eq:s2}
	s \triangleq \sum_{\ell=1}^{m_2} u_2^\ell (v_2^\ell\strut^\TRANS x_0)  
\end{equation}

where $\{u_2^\ell\}_{\ell=1}^d$ (resp. $\{v_2^\ell\strut^\TRANS\}_{\ell=1}^d$) represents the set of right (resp. left) orthonormal eigenvectors of $\bar L$ associated with $\lambda_2(\bar L)$,  and $x_0$ denotes the initial state of opinions.  
\item The signs of elements in $s$ 
separate the nodes in the network into two clusters as follows:
\[
\begin{aligned}
	&C_{1} =\{i:  s(i)>0\},\quad 
	&C_{2} =\{i:  s(i)\leq 0\}.
\end{aligned}
\]

\end{enumerate}

\vspace{5pt}

The features of this partition algorithm for social networks (or social choice algorithm) include: 
\begin{itemize}
 \item When the algebraic multiplicity of $\lambda_2(\bar L)$ is $1$, the graph $\mathcal{G}(A)$ can be partitioned into the same two clusters regardless of the  initial opinion state $x_0$ (if $x_0\neq 0$). 
 \item When the algebraic multiplicity of $\lambda_2(\bar L)$ of $\bar L$ is greater than $1$, we need to take the initial opinion state $x_0$ into consideration when partitioning the network. 
 \item With any non-zero initial condition $x_0$, the algorithm can always  partition the nodes of $\mathcal{G}(A)$ into two clusters since  $\sum_{i=1}^n s(i) =0$ always holds due to the fact that  $u_2^\ell$ is orthogonal to the eigenvector $u_1= \frac{1}{\sqrt{n}}\mathbf{1}_n$ for all  $\ell \in\{1,\ldots, m_2\}$.
\item  The partition depends on the eigen direction associate with $\lambda_2(\bar L)$ under the interpretation that the disagreement state projected into the disagreement subspace associated with $\lambda_2(\bar L)$ will eventually become relatively significant when the time is long enough.  
\item  The partition does not change over time since replacing $x_0$ by $x(t)$ in \eqref{eq:s2} does not change the signs of the elements of $s$.
\end{itemize}

\begin{remark}
	When accurate information on the initial opinion state $x_0$ and the time constant $\tau$ is available, one may characterize the exact evolution of the opinions and hence establish a time-varying partition of the nodes on the graph based on their disagreement states $x_i^{\text{dis}}(t), i\in [n]$. 
\end{remark}
\begin{remark}
 When $\lambda_2(\bar L)$ of $\bar L$ has multiplicity more than one and $x_0$ is not known, then there is no decisive partition.  A trivial example is the complete graph with $n~ $ nodes $(n>2)$ and uniform weights. In this case, the multiplicity of $\lambda_2(\bar L)$ is $n-1$ and the signs of elements in $s$ depends on $x_0$, and hence there is no meaningful partition based on  the graph structure only.	
\end{remark}

\subsection{Applications to real-world social networks}
\subsubsection{Application to Zachary's karate club network}

During Zachary's study of the social structure in a karate club \cite{zachary1977information}, a conflict between the administrator (node 34) and the instructor (node 1) divided the club into two groups. Each node on the network represents an individual person and edges represent their social interactions outside the club.  In \cite{zachary1977information} based on maximum-flow-minimum-cut analysis of the unweighted network structure, all but one member (node 9) of the club were correctly assigned individuals to groups they actually joined after the split.

\begin{figure}[htb]
\centering
\vspace{5pt}
	\includegraphics[frame,width=7cm,trim={4cm 2cm 2cm 2cm},clip]{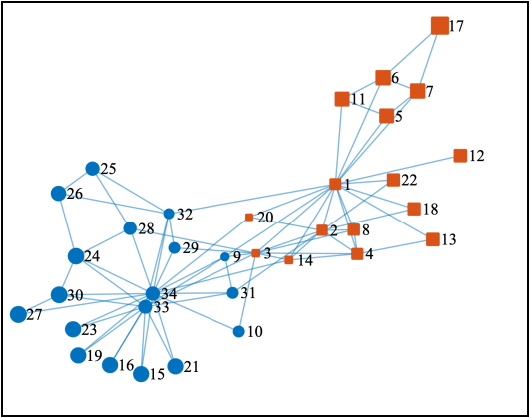}
	\caption{Zachary's karate club network structure and the two clusters following our partition algorithm. Square-shaped nodes belong to cluster $C_1$ and circle-shaped nodes belong to cluster $C_2$. The sizes of the nodes are monotone with respect to the strength of their memberships to their own clusters.}\label{fig:network-split}
\end{figure}
An application of our partition algorithm to the unweighted Karate club network assigns nodes into two separate groups:
\begin{equation}
\begin{aligned}
	C_1 &= \{1,2,3, 4,5,6,7,8,11,12,13,14,17,18,20,22\},\\
	C_2 & = \{9,10,15,16,19,21, 23,24,...,33,34\},
\end{aligned}
	\end{equation}
as illustrated in Figure \ref{fig:network-split}. 	
%
This clustering result coincides exactly with the division in the actual situation  \cite{zachary1977information} and provides the same clustering result based on the modularity approach in \cite{newman2006modularity}. 

 It is worth emphasizing that the Fielder eigenvector of the original social network $\mathcal{G}(A)$ would assign all but node $3$ into the correct groups.  
 %
 This indicates that the degree-centrality-weighted influence matrix $\bar A$ provides a more accurate spectral clustering result than the underlying social network structure represented by $A$. 
 \vspace{3pt}

\subsubsection{Application to the southern woman network}

\begin{figure}[htb]	
\centering
\includegraphics[frame, width=7cm, trim={3cm 2cm  2cm 2cm},clip]{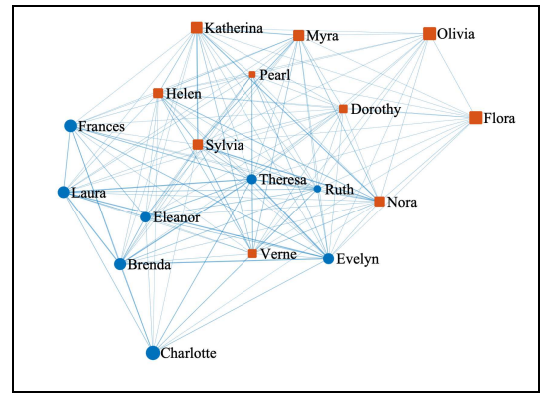}
	\caption{Partition of the southern woman networks into two clusters.}\label{fig:southern-woman-net}
\end{figure}

The southern women network structure is analyzed and  clustered them into groups in \cite{davis1941deep} based on interviews of 18 women. These 18 women attended 14 events and the connections among them are characterized by the number of co-attended events. Our partition algorithms assign individuals to two groups which is the same bipartition result except one node (node Pearl) as those in \cite{davis1941deep} and \cite{liebig2014identifying}. 
The partition result is as follows:
\begin{itemize}
	\item Cluster $C_1$ characterized by square-shaped nodes consists of  Dorothy, Flora, Helen, Katherina, Myra, Nora, Olivia, Sylvia, Verne; 
	\item Cluster $C_2$ characterized by circle-shaped nodes consists of Brenda, Charlotte, Eleanor, Evelyn, Frances, Laura, Pearl, Ruth, Theresa.
\end{itemize}
as illustrated in Figure \ref{fig:southern-woman-net}.

In contrast, the Fielder eigenvector of the original social network $\mathcal{G}(A)$ would partition nodes into two groups where one group only consists of Flora and Olivia, which is far from the real-world clustering result in \cite{davis1941deep}. This again indicates that the degree (centrality) weighted influence is important analyzing social network structures and opinion dynamics. 

\subsection{Network partition into multiple groups}
There are two ways to partition the graph into multiple clusters: 1) iterative bipartition and 2) $K$-means.  These two ways represent slightly different meanings of the partition. 
\vspace{3pt}
\subsubsection{Iterative bipartition}

Without initial opinion states and time constant, the partition of nodes on the network  into different groups can be carried out as follows: 
First, we partition the graph into two subgraphs following our partition algorithm. Then we partition each of the subgraphs via our same partition algorithm. This procedure continues  until the multiplicity of second smallest eigenvalue of the Laplacian for any graph or subgraph is more than $1$.  We then create a partition algorithm that can partition the graph into multiple groups.  One feature of this partition method is that the number of clusters is automatically determined from the partition procedure and there is no need to specify the number of clusters beforehand.
\vspace{3pt}
\subsubsection{K-means}
The vector $s$ produced by our algorithm approximately quantifies the disagreement level of individuals on the network.
If the number of partitions is fixed and known beforehand, we can implement the standard $K$-means \cite{arthur2006k} to cluster the approximate disagreement state values $\{s(i), i\in[n]\}$. We note that when $\lambda_2(\bar L)$ has multiplicity more than 1,  the initial opinion states $x_0$ is also needed.  
 By clustering the disagreement states, we provide a partition of the nodes into different clusters, 
 where different clusters represent nodes with different levels of disagreements. 
Furthermore, if both the time constant $\tau$ and the initial opinion states $x_0$ are known, then one can exactly characterize the evolution of the disagreement state and apply $K$-means to identify time-varying clusters. 

\subsection{Diversity of opinions}

\subsubsection{Opinion diversity energy}
Consider the following the energy function $E(\cdot):\mathbb{R}^n\to [0,\infty)$ with
\[
E(z) := \frac12 \sum_{i=1}^n \sum_{j=1}^n {a}_{ij}(z_i-z_j)^2 = z^\TRANS L z,\quad  z \in \mathbb{R}^n, 
\]
where $L := \textup{diag}(A\mathbf{1})- A$. 
We note that here $a_{ij}$ instead of $\bar a_{ij}$ is used.
%
The \emph{opinion diversity energy} at time $t$ is then given by 
\begin{equation}
	\begin{aligned}
	E(x^\textup{dis}(t)) & = \frac12\sum_{i=1}^n \sum_{j=1}^n {a}_{ij}(x^\textup{dis}_i (t)-x^\textup{dis}_j(t))^2
	\end{aligned} 
\end{equation}
The opinion projected into the eigendirection associated with $\lambda_2(\bar L)$ is given by 
$$
y_2(t):= \sum_{\ell=1}^{m_2} u_2^\ell v_2^\ell\strut^\TRANS x(t) = e^{-\frac{t}{\tau} \lambda_2} \sum_{\ell=1}^{m_2} u_2^\ell v_2^\ell\strut^\TRANS x_0.
$$
and the associated diversity energy is then given by  
\[
E(y_2(t)) = \frac12\sum_{i=1}^n \sum_{j=1}^n {a}_{ij}(y_{2i} (t)-y_{2j}(t))^2 = y_2(t)^\TRANS L y_2(t).
\]
\begin{remark}
	There may be other choices of the energy function $E(\cdot)$ with slightly different meanings,  such as
\[
\begin{aligned}
	&E(z) :=  z^\TRANS \bar L z 
	\quad \text{or}\quad E(z) :=\frac12 \sum_{i=1}^n \sum_{j=1}^n {\bar a}_{ij}(z_i-z_j)^2.
\end{aligned}\]
\end{remark}

\subsubsection{Inverse entropy diversity}
If $\lambda_2(\bar L)$ has multiplicity $1$, we may  use the following approximate estimate of the opinion diversity in \eqref{eq:opp-diveristy}, without considering the underlying time constant $\tau$ and initial condition $x_0$. 
 An entropy associated with $u_2$ can be defined via 
\begin{equation}
	H (u_2) = -\sum_{i=1}^n (u_2(i))^2 \log (u_2(i))^2,
\end{equation}
where $u_2$ is the normalized Fiedler eigenvector (i.e., the normalized eigenvector with $\lambda_2(\bar L)$). 
The diversity of opinions (or the size of disagreement)  on the graph $\mathcal{G}(\bar A)$ can  then be characterized by 
\begin{equation}\label{eq:opp-diveristy}
	D (\mathcal{G}(\bar A)) = H^{-1}(u_2)= \frac{1}{-\sum_{i=1}^n (u_2(i))^2 \log (u_2(i))^2}.
\end{equation}
Since the property $\langle u_2, \mathbf{1}\rangle =0$ holds,
the index  takes into account the signs of $u_2$ implicitly. Roughly speaking, this index measures the diversity of the relative membership strengths of individuals to their own opinion groups.

Following the idea of Inverse Simpson index, another diversity measure can be given by 
\begin{equation}
	D (u_2) = \frac{1}{\sum_{i=1}^n (u_2(i))^4}.
\end{equation}

 \subsection{The Markov chain interpretation}

Following \eqref{eq:property-Abar}, each row of $\bar{A}$ sums up to one and hence $\bar{A}^\TRANS$ can be associated to the probability transition matrix of a Markov chain.  
Let $p_{ki}$ represent the probability of individual $i$ support an idea or an opinion at time $k$.   
Then the probability transition is characterized as follows: 
 $
 p_{(k+1)i} = \sum_{j=1}^n p_{kj} \bar{A}^\TRANS_{ji} = \sum_{j=1}^n p_{kj} \bar{A}_{ij} 
 $
With probability (row) vector  $p_k = [p_{k1},\ldots, p_{kn}]$, the probability transition is compactly characterized by
 \begin{equation} \label{eq:markov-transition}
 p_{k+1} = p_k \bar{A}^\TRANS. 
 \end{equation}
This formulation follows the Degroot model \cite{degroot1974reaching} by specializing the influence matrix in \cite{degroot1974reaching} to be the degree-centrality-weighted matrix $\bar A^\TRANS$ given by
 \[
 \bar{A}_{ij}=\frac{d_j}{\sum_{j\in N_i}a_{ij}d_j} a_{ij},
 \quad i,j \in \{1,...,n\}.
 \]
\begin{proposition}[Appendix \ref{app:prop4}]\label{prop:markov-eigenvalue}
If the underlying graph $\mathcal{G}(A)$ with the adjacency matrix $A$ is connected, then 
the Markov chain associated with $\bar A^\TRANS$ as in \eqref{eq:markov-transition} is irreducible and aperiodic if and only if $-1$ is not an eigenvalue of $\bar A$.  
\end{proposition}

  The diversity of the possible opinions at time $k$ can be estimated based on the inverse of the entropy of $p_k$, that is,  $D_k = {E_k^{-1}}, \text{where}~ 
  E_k = - \sum_{i=1}^n p_{ki} \log p_{ki}.
$
  If $\mathcal{G}(A)$ with the adjacency matrix $A$ is undirected and connected and $\bar A$ does not have eigenvalue $-1$, that is,  the underlying Markov chain is irreducible and aperiodic, then $p_{\infty i}= \frac1n$ for all $i\in [n]$ and $D_\infty  = (\log n)^{-1}$.

\subsection{General centrality-weighted opinion dynamics}

Centralities on a network, which typically depend on the network structures,  quantify the importance of nodes on the network. 
The degree centrality explored in Section II-A is a particular choice of centrality. 
Other examples of centralities included betweenness, eigen-centrality, page-rank centrality, Sharply values, etc.  
For different types of networks, different centralities may be suitable to characterize the influence on  information and opinion propagation. 

 Similar to the previous model in Section II-A, the basic assumptions for general centrality-weighted opinion dynamics include the following:
 \begin{enumerate}
 	\item[(i)] Individual on a social network communicate and change their own opinions in the direction to conform with those of their neighbours;
 	\item[(ii)] Each individual weights these influences from the neighbours based on their importance on the network quantified by the centrality vector $\rho$.
 \end{enumerate}
 The dynamic model for opinion state evolution over a network is then given by
\[
\tau \dot {x}_i =  \sum_{j\in N_i} \frac{\rho_j}{\sum_{j\in N_i}a_{ij}\rho_j} a_{ij}(x_j-x_i)
 \]
 where $\rho_i > 0$ is the centrality of node $i$ on the network (based on an appropriate choice of centrality), $\tau>0$ is an appropriate time constant, and $a_{ij}$ represents the connection from node $j$ to $i$.  The centrality $\rho$ should be chosen according to the underlying application problems, as different centralities may be suitable for different application problems. 
 The Markov chain interpretation  in \eqref{eq:markov-transition} can also be generalized simply by replacing $d_i$ there by $\rho_i$. 

If $\rho_i>0$ for all $i \in [n]$, all the results in \textbf{Propositions 1-4} hold for general centralities as well.  The corresponding partition algorithm extends this this case naturally.  
Furthermore, one can verify that when the centrality $\rho(\cdot)$ is time-varying or state-dependent,  all the results in \textbf{Propositions 1-3} hold as long as $\rho_i(\cdot)>0$ for all $i \in [n]$. %

\section{Conclusions}
This paper proposed an opinion dynamics model based on network structures and nodal centralities. The model was used to partition graphs into clusters. %

Future work should include exploring more real-world examples on different types of social networks, studying similar models for directed graphs, and providing a systematic procedure to identify suitable centralities based on data (i.e., the learning of the suitable centrality on social networks). Moreover, the centrality may also be generalized to depend on the opinion states or some equilibrium states. 

\section{Acknowledgement}
The author would like to thank Prof. Peter Caines for the helpful conversations. 
\appendix[Proofs of Propositions 1-4] 
\subsection{Proof of Proposition \ref{prop:one-zero-eigenvalue}} \label{app:prop1}
\begin{proof}
One can easily verify that $\bar L \mathbf{1} =0$, that is,  $0$ is an eigenvalue of $\bar L$. 
	Based on Gershgorin disk theorem \cite{gersgorin1931uber}, among all points on the imaginary axis only the origin can be the eigenvalue of $\bar  L$, and all the eigenvalues except $0$ have strictly positive real parts.   
	If any directed graph contains a rooted out-branching\footnote{A \emph{rooted out-branching} on a directed graph is defined as the directed subgraph which is a directed tree, consists of all the nodes of the original graph, and contains a single root node (i.e., the node that has a directed path to all other nodes). }, then the rank of the Laplacian matrix is $n-1$ (see for instance \cite{mesbahi2010graph}). 
 In the current case, since $A$ corresponds to an undirected and connected graph,  we obtain that $d_i>0$ for all $i$ and furthermore we note that $\bar{A}$ is the adjacency matrix of a  strongly connected directed graph.  Therefore $\bar{A}$ corresponds to the adjacency matrix of a graph that contains a rooted out-branching. Hence $\bar L$ as the associated Laplacian matrix has only one zero eigenvalue.
\end{proof}

\subsection{Proof of Proposition \ref{prop:realeigenpairs}} \label{app:prop2}

\begin{proof}
	Let $P = \diag^{\frac12}(d) \diag^{\frac12}(h).$ Since for any connected graph all the elements of $d$ and those of $h= Ad$ are non-zero, the inverse of $P$ exists and is given by
	 $P^{-1}= \diag^{-\frac12}(d)\diag^{-\frac12}(h).$
	Multiplying $P$ and $P^{-1}$ on the left and right side of $\bar A$ yields
	\begin{equation}\label{eq:TransAbar}
	\begin{aligned}
			P \bar{A} P^{-1} 
& =  \diag^{\frac12}(d) \diag^{-\frac12}(h)  A 
\diag^{\frac12}(d)\diag^{-\frac12}(h) \\
& \triangleq Q^\TRANS A Q
	\end{aligned}
		\end{equation}
	where $Q = \diag^{\frac12}(d) \diag^{-\frac12}(h)$. Since $A$ is symmetric, $Q^\TRANS A Q$ is also symmetric, and hence all the eigenvalues of  $Q^\TRANS A Q$ are real. As is known that the similarity transformation 
	$\bar{A} = P^{-1} \left(Q^\TRANS A Q\right) P$
preserves all eigenvalues, we obtain that all the eigenvalues of $\bar{A}$ are real. 
	Since $\bar L  = I_n - \bar{A}$, the eigenvalues $\lambda_i(\bar L) = 1- \lambda_i(\bar A)$ for all $i$. Therefore all the eigenvalues of $\bar L$ are also real.  
\end{proof}

\subsection{Proof of Proposition \ref{prop:diagonalizable}} \label{app:prop3}

\begin{proof}
	Recall from \eqref{eq:TransAbar} that
\[
	\begin{aligned}
			P \bar{A} P^{-1} 
& \triangleq Q^\TRANS A Q,\quad  Q = \diag^{\frac12}(d) \diag^{-\frac12}(h)
	\end{aligned}
	\]
and  $ Q^\TRANS A Q$ is symmetric and obviously diagonalizable. Hence  $P \bar{A} P^{-1} $	is diagonalizable.   That is, there exist an invertible matrix $V$ and a diagonal matrix $\Sigma$ such that 
$
 P \bar{A} P^{-1}  = V^{-1} \Sigma  V.
$
Hence $ \bar{A}   = (V P)^{-1} \Sigma  VP$. 
Therefore 
\[
\bar L  =  I_n-\bar A =  (V P)^{-1} (I_n -\Sigma) V P,
\]
that is, $\bar L$ is diagonalizable. 
%
\end{proof}
\subsection{Proof of Proposition \ref{prop:markov-eigenvalue}} \label{app:prop4}
\begin{proof}
Since $\mathcal{G}(A)$ is connected, $\mathcal{G}(\bar A^\TRANS)$ is strongly connected and hence the Markov chain associated with  $\bar A^\TRANS$ is irreducible. The aperiodicity is equivalent the fact that  $\bar A^\TRANS$ has only one eigenvalue that lies on the unit circle of the complex plane. 
  Since all the eigenvalues of $\bar A$ are real and $1$ is always an eigenvector of $\bar A$, the aperiodicity is equivalent to that $-1$ is not an eigenvalue of $\bar A$. 
\end{proof}
\bibliographystyle{IEEEtran}
\bibliography{mybib}
\end{document}